\documentstyle[prd,aps]{revtex}

\newcommand{\kl}[3]{\mbox{$\rm #1$}^{\mu\nu , \alpha\beta}_{#2}(#3)}

\begin{document}
\draft
\title{NON-EQUILIBRIUM DYNAMICS OF A THERMAL PLASMA
       IN A GRAVITATIONAL FIELD}
\author{Antonio Campos}
\address{Department of Physics,
         University of Maryland,
         College Park, Maryland 20742;
         \\
         Institut de F{\'\i}sica d'Altes Energies,
         Universitat Aut{\`o}noma de Barcelona,
         08193 Bellaterra (Barcelona), Spain;
         \\
         Center for Theoretical Physics, 
         Laboratory for Nuclear Science,
         Massachusetts Institute of Technology, 
         Cambridge, Massachusetts 02139\footnote{Present address}}
\author{B. L. Hu}
\address{Department of Physics,
         University of Maryland,
         College Park, Maryland 20742}
\author{Report umdpp \#98-98/MIT-CTP 2745}\date{\today}
\maketitle
\begin{abstract}
{\scriptsize
We introduce functional methods to study the non-equilibrium dynamics
of a quantum massless scalar field
at finite temperature in a gravitational field.
We calculate the close time path (CTP) effective action and, using its
formal equivalence with the influence functional, derive
the noise and dissipation kernels of the quantum open system
in terms of quantities in thermodynamical equilibrium.
Using this fact, we formally
prove the existence of a fluctuation-dissipation relation (FDR) at all
temperatures between the quantum fluctuations
of the plasma in thermal equilibrium and the energy dissipated by 
the external gravitational field. What is new is the identification of
a stochastic source (noise) term arising from the quantum and thermal
fluctuations in the plasma field, and the derivation of
a Langevin-type equation which describes the non-equilibrium dynamics
of the gravitational field influenced by the plasma. The back reaction
of the plasma on the gravitational field is embodied in the FDR.
From the CTP effective action
the contribution of the quantum scalar field to the thermal 
graviton polarization tensor can also be derived and it is shown
to agree with other techniques, most notably, linear response theory
(LRT). We show the connection between the LRT, which is
applicable for near-equilibrium conditions and the functional
methods used in this work which are useful for fully non-equilibrium 
conditions.}
\end{abstract}
\baselineskip=15pt


\section{Introduction}
\label{sec:intro}


The behavior of a relativistic quantum field at finite temperature
in a gravitational field has been studied before by Gross, Perry 
and Yaffe \cite{GPY82}, Gribosky, Donoghue and Holstein \cite{GDH89},
Rebhan and co-workers \cite{Reb91}, de Almeida, Brandt, Frenkel and 
Taylor \cite{ABF94,Taylor} for scalar and Abelian gauge fields.
The thermal graviton polarization tensor and the effective action
have been calculated and applied to the study of stability of hot
flat/curved spaces, and dynamics of cosmological perturbations. Using 
linear response theory (LRT) \cite{Ein05,Nyq28,CW51,Kub57,KMS,FDR}, 
Jackiw and Nair \cite{JN93} derived the response function at high 
temperatures for a non-Abelian gauge field in a thermal plasma of 
charged particles. 

To describe screening effects and stability of thermal quantum 
gravity, one needs only the real part of the polarization tensor; 
but for damping effects, the imaginary part is essential. The 
gravitational polarization tensor obtained from the thermal 
graviton self-energy represents only a part (the thermal
correction to the vacuum polarization) of the finite temperature
quantum stress tensor, which under more general conditions
(e. g. dynamical background) should contain also contributions 
from particle creation (arising from vacuum fluctuations at zero and 
finite temperature). These processes engender dissipation in the 
dynamics of the gravitational field and their fluctuations appear 
as noise in the plasma. These latter two effects have not been 
sufficiently appreciated in most previous works in thermal field
theory.

In this work we use open system concepts and functional methods a la 
Schwinger-Keldysh \cite{Sch61,Kel65} and Feynman-Vernon \cite{FV63}. 
This framework encompasses LRT which we illustrate in a calculation 
of the quantum corrections of the scalar field to the thermal graviton
polarization tensor with results identical to that obtained before 
\cite{Reb91,ABF94}. More importantly, we derive the noise and 
dissipation kernels, prove that they satisfy a 
fluctuation-dissipation relation (FDR) \cite{CW51,Kub57,FDR} at all 
temperatures, and derive a stochastic semiclassical equation for the 
non-equilibrium dynamics of the gravitational field under the
influence of the thermal plasma including all three aspects mentioned
above. These cannot be obtained easily by LRT and other thermal field 
theory methods.

Here the thermal plasma will be described by a massless scalar quantum
field at finite temperature and the gravitational field by a linear
perturbation from Minkowski spacetime. Since the plasma is considered 
to be in thermal equilibrium it will be characterized by a time-like 
normalized vector field $u^\mu$, representing the four-velocity of 
the plasma, and the temperature of the fluid in its own rest frame 
$\beta^{-1}$. Taking into account the four-velocity $u^\mu$ of the 
plasma, a manifestly Lorentz-covariant approach to thermal field 
theory may be used \cite{Wel82}. However, in order to simplify the 
involved tensorial structure we work in the co-moving coordinate 
system of the plasma where $u^\mu = (1,0,0,0)$.

By making conformal transformations on the field and the spacetime,
our results may be easily generalized to the case of a conformally 
coupled quantum scalar field at finite temperature in a spatially 
flat Friedmann-Robertson-Walker universe \cite{Hu82}. Indeed we 
have earlier used the functional method and the Brownian motion 
paradigm \cite{Sch61,FV63,CL83,HPZ} to study similar problems in 
semiclassical gravity \cite{BD82}. We found that quantum noise
arising from fluctuations in the particle creation would constitute
a stochastic source hitherto undiscovered \cite{Hu89} (whose effect 
can overdominate the expectation value of the energy momentum tensor
in the semiclassical Einstein equation \cite{KF93,FW96,PH97})
in a new form of Einstein-Langevin equation 
\cite{CH94,HM95,HS95,CV96,LM97,CV97,CCV97,Hal97,MV97}. We also came 
to the understanding that back reaction of vacuum quantum field 
processes (such as particle creation) on the dynamics of the early
universe near the Planck time is summarily a manifestation of a FDR 
in semiclassical gravity.

The  close time path (CTP) functional formalism of Schwinger  
\cite{Sch61} is based on a generating functional defined in 
such a way that expectation values (rather than matrix elements) 
of time and anti-time ordered field products are directly computed. 
Keldysh \cite{Kel65} developed a diagramatic technique approach 
useful for non-equilibrium processes in quantum field theory. This 
method has been further developed in \cite{CSH80,CSHY85,CCSY88} 
(For some applications in field theory and cosmology see 
\cite{CH87,CH88,BL93,BDH94,Hu94,CH94,CV94,GR94,CHKMP94,CH95,RH97}). 
The advantage of this formalism is that effective dynamical equations 
for the expectation value of the quantum field may be derived from 
the effective action. Since these effective equations are real and 
causal a proper initial value problem may be formulated. The 
generalization of this formalism to curved spacetimes was given by 
DeWitt and Jordan, Calzetta and Hu \cite{DeW86,Jor86,CH87}.

In Sec.~\ref{sec:effective action}, we describe our model and 
the derivation of the thermal CTP effective action. From this
effective action we identify the dissipation and the noise kernels
representing the linear response and the quantum fluctuations of the
plasma, respectively, and obtain a stochastic semiclassical equation
for the gravitational perturbations. This is done by using the
underlying ideas of the influence functional formalism of Feynman and 
Vernon \cite{FV63}. Sec.~\ref{sec:FDR} is devoted to showing the 
existence of the FDR. We formally show that the noise and dissipation 
kernels extracted from the CTP effective action in general satisfy a 
formal relation at all temperatures. By explicitly computing the 
corresponding kernels we show that this relation is satisfied at high 
and zero temperature. In Sec.~\ref{sec:LRT} we describe how our 
functional approach reproduces results from LRT. Finally, in 
Sec.~\ref{sec:conclusions}, we give a brief conclusion. To make 
this presentation self-contained,  in Appendix~\ref{app:TFT} we give 
a brief yet complete description of thermal field theory via the CTP 
functional formalism.


\section{CTP effective action at finite temperature}
\label{sec:effective action}


\subsection{The model}
 
In this section, we derive the CTP effective action for a thermal
plasma in a gravitational field. To describe the plasma we consider a
free massless scalar field $\phi$ arbitrarily coupled to a
gravitational field $g_{\mu\nu}$ with classical action
\begin{equation}
   S_m[\phi,g_{\mu\nu}]
        \ = \ -{1\over2}\int d^nx\ \sqrt{-g}
               \left[ g^{\mu\nu}\partial_\mu\phi\partial_\nu\phi
                     +\xi(n) R\phi^2
               \right],
\end{equation}
where $R(x)$ is the scalar curvature and the arbitrary parameter 
$\xi(n)$ defines the type of coupling between the scalar field and the
gravitational field. If $\xi(n) = 0$ the quantum field is said to be
minimally coupled, if $\xi(n) = (n-2)/[4(n-1)]$, where $n$ is the
spacetime dimensions, the field is said to be conformally coupled. To
describe the gravitational field we consider a small deviation from
flat spacetime
\begin{equation}
   g_{\mu\nu}(x) 
        \ = \ \eta_{\mu\nu} + h_{\mu\nu}(x), 
\end{equation}
with signature $(-,+,\cdots ,+)$ for the Minkowski metric. 
Using this metric and neglecting the surface terms that appear
in an integration by parts, the action for the scalar field may be 
written perturbatively as 
\begin{equation}
   S_m[\phi,h_{\mu\nu}]
        \ = \  {1\over2}\int d^nx\ \phi
               \left[ \Box + V^{(1)} + V^{(2)} + \cdots
               \right] \phi,
\end{equation}
where the first and second order perturbative operators $V^{(1)}$ and 
$V^{(2)}$ are given by 
\begin{eqnarray}
   V^{(1)}
        & \ \equiv \ & - \left\{ [\partial_\mu\bar h^{\mu\nu}(x)]
                                 \partial_\nu
                                +\bar h^{\mu\nu}(x)\partial_\mu
                                 \partial_\nu
                                +\xi(n) R^{(1)}(x)
                          \right\},
                     \nonumber \\
   V^{(2)}
        & \ \equiv \ & \left\{ [\partial_\mu\hat h^{\mu\nu}(x)]
                               \partial_\nu
                              +\hat h^{\mu\nu}(x)\partial_\mu
                               \partial_\nu
                              -\xi(n) ( R^{(2)}(x)
                                       +{1\over2}h(x)R^{(1)}(x))
                       \right\}.   
\end{eqnarray}
In the above expressions, $R^{(k)}$ is the $k$-order term in the
pertubation $h_{\mu\nu}(x)$ of the scalar curvature and the
definitions $\bar h_{\mu\nu}$ and $\hat h_{\mu\nu}$ denote a
linear and a quadratic combination of the perturbation, respectively,
\begin{eqnarray}
   \bar h_{\mu\nu}
        & \ \equiv \ & h_{\mu\nu} - {1\over2} h \eta_{\mu\nu},
                     \nonumber \\
   \hat h_{\mu\nu}
        & \ \equiv \ & h^{\,\, \alpha}_\mu h_{\alpha\nu}
                      -{1\over2} h h_{\mu\nu}
                      +{1\over8} h^2 \eta_{\mu\nu}
                      -{1\over4} h_{\alpha\beta}h^{\alpha\beta} 
                       \eta_{\mu\nu}.
   \label{eq:def bar h}
\end{eqnarray}

For the gravitational field we take the following action
\begin{eqnarray}
   S^{div}_g[g_{\mu\nu}]
        & \ = \ & {1\over\ell^{n-2}_P}\int d^nx\ \sqrt{-g}R(x)
                \nonumber \\
        &       & +{\alpha\bar\mu^{n-4}\over4(n-4)}
                   \int d^nx\ \sqrt{-g}
                   \left[ 3R_{\mu\nu\alpha\beta}(x)
                           R^{\mu\nu\alpha\beta}(x)
                         -\left( 1-360(\xi(n) - {1\over6})^2
                          \right)R(x)R(x)
                   \right].
\end{eqnarray}
The first term is the classical Einstein-Hilbert action and the second
divergent term in four dimensions is the counterterm used in order to 
renormalize the effective action. In this action $\ell^2_P = 16\pi G$,
$\alpha = (2880\pi^2)^{-1}$ and $\bar\mu$ is an arbitrary mass scale.
It is noteworthy that the counterterms are independent of the
temperature because the thermal contribution to the effective action
is finite and does not include additional divergencies.

\subsection{CTP effective action}
\label{subsec:CTP eff act}

In Appendix~\ref{app:TFT} we described how to compute the CTP
effective action at finite temperature for a quantum scalar free field
theory. In fact, we need to generalize the CTP formalism for two
fields in the semiclassical approximation. This generalization, which
is justified in Appendix~\ref{app:just}, gives the following formal 
effective action (replace $\phi$ by $h_{\mu\nu}$ and $\psi$ by $\phi$ 
in equation (\ref{eq:twofields}))
\begin{equation}
   \Gamma^\beta_{CTP}[h^\pm_{\mu\nu}]
        \ = \ S^{div}_g[h^+_{\mu\nu}] 
             -S^{div}_g[h^-_{\mu\nu}]
             -{i\over2}Tr\{ \ln\bar G^\beta_{ab}[h^\pm_{\mu\nu}]\},
   \label{eq:eff act two fields}       
\end{equation}
where $\pm$ denote the forward and backward time path of the CTP
formalism and $\bar G^\beta_{ab}[h^\pm_{\mu\nu}]$ is the complete 
$2\times 2$ matrix propagator ($a$ and $b$ take $\pm$ values) with 
thermal boundary conditions for the differential operator 
$\Box + V^{(1)} + V^{(2)} + \cdots$. In fact, the actual form of 
$\bar G^\beta_{ab}$ cannot be explicitly given. However, it is easy 
to obtain a perturbative expansion in terms of $V^{(k)}_{ab}$, the 
$k$-order matrix version of the complete differential operator 
defined by $V^{(k)}_{\pm\pm} \equiv \pm V^{(k)}_{\pm}$ and 
$V^{(k)}_{\pm\mp} \equiv 0$, and $G^\beta_{ab}$, the thermal matrix 
propagator for a massless scalar field in flat spacetime 
(\ref{eq:thermal prop 1}-\ref{eq:thermal prop 2}).
To second order $\bar G^\beta_{ab}$ reads,
\begin{eqnarray}
   \bar G^\beta_{ab}
        \ = \  G^\beta_{ab}
              -G^\beta_{ac}V^{(1)}_{cd}G^\beta_{db}
              -G^\beta_{ac}V^{(2)}_{cd}G^\beta_{db}
              +G^\beta_{ac}V^{(1)}_{cd}G^\beta_{de}
               V^{(1)}_{ef}G^\beta_{fb}
              +\cdots
\end{eqnarray}
Expanding the logarithm and dropping one term independent of the
perturbation $h^\pm_{\mu\nu}(x)$, the CTP effective action may be
perturbatively written as,
\begin{eqnarray}
   \Gamma^\beta_{CTP}[h^\pm_{\mu\nu}]
        & \ = \ &  S^{div}_g[h^+_{\mu\nu}] - S^{div}_g[h^-_{\mu\nu}]
                \nonumber \\
        &       & +{i\over2}Tr[ V^{(1)}_{+}G^\beta_{++}
                               -V^{(1)}_{-}G^\beta_{--}
                               +V^{(2)}_{+}G^\beta_{++}
                               -V^{(2)}_{-}G^\beta_{--}
                              ]
                \nonumber \\
        &       & -{i\over4}Tr[  V^{(1)}_{+}G^\beta_{++}
                                 V^{(1)}_{+}G^\beta_{++}
                               + V^{(1)}_{-}G^\beta_{--}
                                 V^{(1)}_{-}G^\beta_{--}
                               -2V^{(1)}_{+}G^\beta_{+-}
                                 V^{(1)}_{-}G^\beta_{-+}
                              ]. 
   \label{eq:effective action}
\end{eqnarray}

The next task is to identify the noise and dissipation kernels from
the non-local terms of the above thermal CTP effective action. The
last trace is the only term that is responsible for dissipation
because all others are constructed with symmetric kernels. 
In fact, this non-local term have information of the noise kernel also,
as we will see in the next section when we show the existence of a FDR.
In order to identify the role of each kernel we have to compute all
the traces in the formal expression for the effective action.
Some of these traces have divergencies that are canceled using the
counterterms introduced in the classical gravitational action after
dimensional regularization. In general, the non-local pieces are of the
form $Tr[V^{(1)}_{a}G^\beta_{mn}V^{(1)}_{b}G^\beta_{rs}]$.
In terms of the Fourier transformed thermal propagators 
$\tilde G^\beta_{ab}(k)$ these traces can be written as,
\begin{equation}
   Tr[V^{(1)}_{a}G^\beta_{mn}V^{(1)}_{b}G^\beta_{rs}]
        \ = \  \int d^nxd^nx'\ 
               h^a_{\mu\nu}(x)h^b_{\alpha\beta}(x')
               \int {d^nk\over(2\pi)^n}{d^nq\over(2\pi)^n}
               e^{ik\cdot (x-x')}
               \tilde G^\beta_{mn}(k+q)\tilde G^\beta_{rs}(q)
               \kl{T}{}{q,k},
   \label{eq:trace}
\end{equation}
where the tensor $\kl{T}{}{q,k}$ is defined in Appendix~\ref{app:basis}.
In particular, the last trace of (\ref{eq:effective action}) may be 
split in two different kernels $\kl{N}{}{x-x'}$ and $\kl{D}{}{x-x'}$,
\begin{equation}
   {i\over2}Tr[V^{(1)}_{+}G^\beta_{+-}V^{(1)}_{-}G^\beta_{-+}]
        \ = \ -\int d^4xd^4x'\ 
               h^+_{\mu\nu}(x)h^-_{\alpha\beta}(x')
               [   \kl{D}{}{x-x'}
                +i \kl{N}{}{x-x'}
               ].
\end{equation}
Recalling Eq.~(\ref{eq:thermal prop 2}) one can express the Fourier 
transforms of these kernels, respectively, as
\begin{eqnarray}
   \kl{\tilde N}{}{k}
        & \ = \ & \pi^2\int {d^4q\over(2\pi)^4}\ 
                  \left\{ \theta(k^o+q^o)\theta(-q^o)
                         +\theta(-k^o-q^o)\theta(q^o)
                         +n_\beta(|q^o|)+n_\beta(|k^o+q^o|)
                  \right.
                \nonumber \\
        &       & \hskip2cm
                  \left. +2n_\beta(|q^o|)n_\beta(|k^o+q^o|)
                  \right\}\delta(q^2)\delta[(k+q)^2]\kl{T}{}{q,k},
   \label{eq:N}
\end{eqnarray}
\begin{eqnarray}
   \kl{\tilde D}{}{k}
        & \ = \ & -i\pi^2\int {d^4q\over(2\pi)^4}\
                  \left\{ \theta(k^o+q^o)\theta(-q^o)
                         -\theta(-k^o-q^o)\theta(q^o)
                         +sg(k^o+q^o) n_\beta(|q^o|)
                  \right.
                \nonumber \\
        &       & \hskip2cm
                  \left. -sg(q^o)n_\beta(|k^o+q^o|)
                  \right\}\delta(q^2)\delta[(k+q)^2]\kl{T}{}{q,k}.
   \label{eq:D}
\end{eqnarray}
Using the property $\kl{T}{}{q,k} = \kl{T}{}{-q,-k}$, it is easy to 
see that $\kl{N}{}{x-x'}$ is symmetric and $\kl{D}{}{x-x'}$ 
antisymmetric in their arguments; that is, $\kl{N}{}{x} = \kl{N}{}{-x}$ 
and $\kl{D}{}{x} = -\kl{D}{}{-x}$. To properly identify these kernels
with the noise and the dissipation of the system, respectively, we
have to write the renormalized CTP effective action at finite
temperature (\ref{eq:effective action}) in an influence functional form
\cite{FV63}. The imaginary part of the CTP effective action written 
in this form can be identified with the noise kernel and the 
antisymmetric piece of the real part with the dissipation kernel. This
identification will be finally justified in two ways. In
Sec.~\ref{sec:FDR} we will see that these kernels satisfy a thermal
FDR and in Sec.~\ref{sec:LRT} we will see from the point of view of
the LRT that $\kl{D}{}{x}$ corresponds to the 
dissipation of the gravitational field or equivalently to the response
of the thermal plasma to the gravitational perturbation, and
$\kl{N}{}{x}$ to the random fluctuations of the plasma. 

If we denote the difference and the sum of the perturbations
$h^\pm_{\mu\nu}$ defined along each branch $C_\pm$ of the complex time
path of integration $C$ by
$[h_{\mu\nu}] \equiv h^+_{\mu\nu} - h^-_{\mu\nu}$ and
$\{h_{\mu\nu}\} \equiv h^+_{\mu\nu} + h^-_{\mu\nu}$, respectively, 
the influence functional form of the thermal CTP effective action may 
be written to second order in $h_{\mu\nu}$ as,
\begin{eqnarray}
   \Gamma^\beta_{CTP}[h^\pm_{\mu\nu}]
        & \ \simeq \ & {1\over2\ell^2_P}\int d^4x\ d^4x'\
                       [h_{\mu\nu}](x)\kl{L}{(o)}{x-x'}
                       \{h_{\alpha\beta}\}(x')
                     \nonumber \\
        &            &+{1\over2}\int d^4x\ 
                       [h_{\mu\nu}](x)T^{\mu\nu}_{(\beta)}
                     \nonumber \\
        &            &+{1\over2}\int d^4x\ d^4x'\ 
                       [h_{\mu\nu}](x)\kl{H}{}{x-x'}
                       \{h_{\alpha\beta}\}(x')
                     \nonumber \\
        &            &-{1\over2}\int d^4x\ d^4x'\ 
                       [h_{\mu\nu}](x)\kl{D}{}{x-x'}
                       \{h_{\alpha\beta}\}(x')
                     \nonumber \\
        &            &+{i\over2}\int d^4x\ d^4x'\ 
                       [h_{\mu\nu}](x)\kl{N}{}{x-x'}
                       [h_{\alpha\beta}](x').
\end{eqnarray}
The first line is the Einstein-Hilbert
action to second order in the perturbation $h^\pm_{\mu\nu}(x)$. 
$\kl{L}{(o)}{x}$ is a symmetric kernel ({\sl i.e.} 
$\kl{L}{(o)}{x}$ = $\kl{L}{(o)}{-x}$) and its Fourier transform is
given by 
\begin{equation}
   \kl{\tilde L}{(o)}{k}
        \ = \ {1\over4}\left[ - k^2 \kl{T}{1}{q,k}
                              +2k^2 \kl{T}{4}{q,k}
                              + \kl{T}{8}{q,k}
                              -2\kl{T}{13}{q,k}
                       \right].
\end{equation}
The fourteen elements of the tensor basis $\kl{T}{i}{q,k}$ 
($i=1,\cdots,14$) are defined in Appendix~\ref{app:basis}. In the
second line $T^{\mu\nu}_{(\beta)}$ has the form of a perfect fluid 
stress-energy tensor
\begin{equation}
   T^{\mu\nu}_{(\beta)}
        \ = \ {\pi^2\over30\beta^4}
              \left[ u^\mu u^\nu + {1\over3}(\eta^{\mu\nu}+u^\mu u^\nu)
              \right],
\end{equation}
where $u^\mu$ is the four-velocity of the plasma and the factor
${\pi^2\over30\beta^4}$ is the familiar thermal energy density for
massless scalar particles at temperature $\beta^{-1}$. In the third
line, the Fourier transform of the symmetric kernel $\kl{H}{}{x}$ can 
be expressed as
\begin{eqnarray}
   \kl{\tilde H}{}{k}
        & \ = \ &  -{\alpha k^4\over4}
                   \left\{ {1\over2}\ln {|k^2|\over\mu^2}\kl{Q}{}{k}
                          +{1\over3}\kl{\bar Q}{}{k}
                   \right\}
                \nonumber \\
        &       &  +{\pi^2\over180\beta^4}
                   \left\{ - \kl{T}{1}{u,k}
                           -2\kl{T}{2}{u,k}
                           + \kl{T}{4}{u,k}
                           +2\kl{T}{5}{u,k}
                   \right\}
                \nonumber \\
        &       &  +{\xi\over96\beta^2}
                   \left\{    k^2 \kl{T}{1}{u,k}
                           -2 k^2 \kl{T}{4}{u,k}
                           -      \kl{T}{8}{u,k}
                           +2     \kl{T}{13}{u,k}
                   \right\}
                \nonumber \\
        &       &  +\pi\int {d^4q\over(2\pi)^4}\
                   \left\{ \delta(q^2)n_\beta(|q^o|)
                           {\cal P}\left[ {1\over(k+q)^2}
                                   \right]
                          +\delta[(k+q)^2]n_\beta(|k^o+q^o|)
                           {\cal P}\left[ {1\over q^2}
                                   \right]
                   \right\}\kl{T}{}{q,k},
   \label{eq:grav pol tensor}
\end{eqnarray}
where $\mu$ is a simple redefinition of the renormalization parameter
$\bar\mu$ given by
$\mu \equiv \bar\mu \exp ({23\over15} +
{1\over2}\ln 4\pi - {1\over2}\gamma)$, and the tensors $\kl{Q}{}{k}$ 
and $\kl{\bar Q}{}{k}$ are defined, respectively, by
\begin{eqnarray}
   \kl{Q}{}{k}
        & \ = \ & {3\over2} \left\{               \kl{T}{1}{q,k}
                                    -{1\over k^2} \kl{T}{8}{q,k}
                                    +{2\over k^4} \kl{T}{12}{q,k}
                            \right\}
                \nonumber \\
        &       &-[1-360(\xi-{1\over6})^2]
                  \left\{               \kl{T}{4}{q,k}
                          +{1\over k^4} \kl{T}{12}{q,k}
                          -{1\over k^2} \kl{T}{13}{q,k}
                  \right\},
   \label{eq:Q tensor}
\end{eqnarray}
\begin{equation}
   \kl{\bar Q}{}{k}
        \ = \  [1+576(\xi-{1\over6})^2-60(\xi-{1\over6})(1-36\xi')]
                  \left\{               \kl{T}{4}{q,k}
                          +{1\over k^4} \kl{T}{12}{q,k}
                          -{1\over k^2} \kl{T}{13}{q,k}
                  \right\}. 
\end{equation}
In the above and subsequent equations, we denote the coupling
parameter in four dimensions $\xi(4)$ by $\xi$ and consequently 
$\xi'$ means $d\xi(n)/dn$ evaluated at $n=4$. $\kl{\tilde H}{}{k}$ 
is the complete contribution of a free massless quantum scalar field 
to the thermal graviton polarization tensor\cite{Reb91,ABF94} and it 
is responsible for the instabilities found in flat spacetime at 
finite temperature \cite{GPY82,Reb91,ABF94}. 
Eq.~(\ref{eq:grav pol tensor}) reflects the fact that the kernel 
$\kl{\tilde H}{}{k}$ has thermal as well as non-thermal
contributions. Note that it reduces to the first term in the zero
temperature limit ($\beta\rightarrow\infty$)
\begin{equation}
   \kl{\tilde H}{}{k}
        \ \simeq \ -{\alpha k^4\over4}
                     \left\{ {1\over2}\ln {|k^2|\over\mu^2}\kl{Q}{}{k}
                            +{1\over3}\kl{\bar Q}{}{k}
                     \right\}.
\end{equation}
This asymptotic contribution to the CTP effective action includes both
pure vacuum quantum fluctuations \cite{CV96} and non-conformal
fluctuations \cite{CV97,CCV97}. On the other hand, the leading term 
($\beta^{-4}$) at high temperature may be written as 
\begin{equation}
   \kl{\tilde H}{}{k}
        \ \simeq \ {\pi^2\over30\beta^4}
                    \sum^{14}_{i=1} 
                    \mbox{\rm H}_i(r) \kl{T}{i}{u,K},
\end{equation}
where we have introduced the dimensionless external momentum
$K^\mu \equiv k^\mu/|\vec{k}| \equiv (r,\hat k)$. The
$\mbox{\rm H}_i(r)$ coefficients were first given in \cite{Reb91} and
generalized to the next-to-leading order ($\beta^{-2}$) in
\cite{ABF94}. They have been reproduced with our sign conventions in 
Appendix~\ref{app:coef} for completeness. It is important to note that
the addition of the contribution of other kinds of matter fields,
even graviton contributions, does not change the tensor structure of 
this leading contribution and only the overall factors are different 
\cite{Reb91}. However, this is not true for the subleading
contributions as was shown in \cite{ABF94}. 

Finally, the noise kernel $\kl{N}{}{x}$ and the dissipation 
kernel $\kl{D}{}{x}$ are new and may be directly read off from
(\ref{eq:N}) and (\ref{eq:D}), respectively. As expected the noise
kernel appears in the imaginary part of the influence functional form
of the CTP effective action and the dissipation kernel is the only 
antisymmetric kernel of the real part. Identification with the
influence functional \cite{FV63} provides a physical interpretation:
$\kl{N}{}{x}$ represents the random fluctuations of the plasma and 
$\kl{D}{}{x}$ the dissipation of energy of the gravitational field.

\subsection{Einstein-Langevin equation}

In this section we show how  a semiclassical
Einstein-Langevin equation  can be derived from the previous
thermal CTP effective action. This equation depicts the stochastic
evolution of the gravitational field under the influence of the
fluctuations of the thermal plasma.

We first introduce the influence functional
${\cal F} \equiv \exp (iS_{IF})$ which is defined from the influence
action $S_{IF}$ of Feynman and Vernon \cite{FV63}. In fact, we can
write this influence functional in terms of the the CTP effective
action because in the semiclassical limit this effective action  
is completely equivalent to $S_{IF}$ \cite{FV63,CCSY88,CH94},
\begin{equation}
   {\cal F}
        \ = \ \exp i\left( Re \{ \Gamma^\beta_{CTP}[h^\pm_{\mu\nu}] \}
                          +{i\over2}\int d^4x\ d^4x'\ 
                               [h_{\mu\nu}](x)\kl{N}{}{x-x'}
                               [h_{\alpha\beta}](x')
                    \right),
\end{equation}
where $Re\{\ \}$ denotes taking the real part. Following
\cite{FV63,CH94,HM95,HS95,CV96}, we can interpret the real part of
the influence functional as the characteristic functional of a
non-dynamical stochastic variable $j^{\mu\nu}(x)$,
\begin{equation}
   \Phi([h_{\mu\nu}])
        \ = \ \exp \left( -{1\over2}\int d^4x\ d^4x'\ 
                           [h_{\mu\nu}](x)\kl{N}{}{x-x'}
                           [h_{\alpha\beta}](x')
                   \right).
   \label{eq:cf}
\end{equation} 
This classical stochastic field represents probabilistically the quantum
fluctuations of the matter field and is responsible for the
dissipation of the gravitational field. By definition, the
above characteristic functional is the functional Fourier transform of
the probability distribution functional ${\cal P}[j^{\mu\nu}]$ with
respect to $j^{\mu\nu}$,
\begin{equation}
   \Phi([h_{\mu\nu}])
        \ = \ \int {\cal D}j^{\mu\nu}\ {\cal P}[j^{\mu\nu}]\
              e^{i\int d^4x\ [h_{\mu\nu}](x)j^{\mu\nu}(x) }.   
   \label{eq:cf_pdf}
\end{equation}
Using (\ref{eq:cf}) one can easily see that the probability
distribution functional is related with the noise kernel by the formal
expression,
\begin{equation}
   {\cal P}[j^{\mu\nu}]
        \ = \ {\exp \left( -{1\over2}\int d^4x\ d^4x'\ 
                           j_{\mu\nu}(x)[\kl{N}{}{x-x'}]^{-1}
                           j_{\alpha\beta}(x')
                   \right)
               \over
               \int {\cal D}j^{\mu\nu}\
               \exp \left( -{1\over2}\int d^4x\ d^4x'\ 
                           j_{\mu\nu}(x)[\kl{N}{}{x-x'}]^{-1}
                           j_{\alpha\beta}(x')
                   \right)
              }.
   \label{eq:pdf}
\end{equation}
For an arbitrary functional of the stochastic field ${\cal A}[j^{\mu\nu}]$,  
the average value with respect to the previous probability distribution
functional is defined as the functional integral 
$\langle {\cal A}[j^{\mu\nu}] \rangle_j 
 \equiv \int {\cal D}[j^{\mu\nu}]\ 
             {\cal P}[j^{\mu\nu}] 
             {\cal A}[j^{\mu\nu}]$.
In terms of this stochastic average the influence functional can be
written as 
${\cal F} = 
 \langle \exp\left(i\Gamma^{st}_{CTP}[h^\pm_{\mu\nu}]
             \right) 
 \rangle_j$, where $\Gamma^{st}_{CTP}[h^\pm_{\mu\nu}]$ is the modified
effective action
\begin{equation}
   \Gamma^{st}_{CTP}[h^\pm_{\mu\nu}]
        \ \equiv \  Re \{ \Gamma^\beta_{CTP}[h^\pm_{\mu\nu}] \}
                   +\int d^4x\ [h_{\mu\nu}](x) j^{\mu\nu}(x).
   \label{eq:mea}
\end{equation}
Clearly, because of the quadratic definition of the characteristic 
functional (\ref{eq:cf}) and its relation with the probability 
distribution functional (\ref{eq:cf_pdf}), the field $j^{\mu\nu}(x)$ is a
zero mean Gaussian stochastic variable. This means that its two-point
correlation function, which is given in terms of the noise kernel by 
\begin{equation}
   \langle j^{\mu\nu}(x) j^{\alpha\beta}(x') \rangle_j
        \ = \ \kl{N}{}{x-x'},
   \label{eq:correlation}
\end{equation}
completely characterizes the stochastic process. The Einstein-Langevin
equation follows from taking the functional derivative of the
stochastic effective action (\ref{eq:mea}) with respect to 
$[h_{\mu\nu}](x)$ and  imposing $[h_{\mu\nu}](x) = 0$ 
\cite{CSH80,CSHY85,CH94}. In our case, this leads to
\begin{equation}
   {1\over\ell^2_P} 
   \int d^4x'\ \kl{L}{(o)}{x-x'} h_{\alpha\beta}(x') 
  +{1\over2}\ T^{\mu\nu}_{(\beta)}
  +\int d^4x'\ \left( \kl{H}{}{x-x'}
                     -\kl{D}{}{x-x'}
               \right) h_{\alpha\beta}(x')
  +j^{\mu\nu}(x)
        \ = \ 0.      
\end{equation}
To obtain a simpler and clearer expression we can rewrite this
stochastic equation for the gravitational perturbation in the harmonic
gauge $\bar h^{\mu\nu}_{\,\,\,\,\, ,\nu} = 0$,
\begin{equation}
   \Box\bar h^{\mu\nu}(x)
         + \ell^2_P
               \left\{ T^{\mu\nu}_{(\beta)}
                      +2P_{\rho\sigma,\alpha\beta}
                       \int d^4x'\ \left( \kl{H}{}{x-x'}
                                         -\kl{D}{}{x-x'}
                                   \right)\bar h^{\rho\sigma}(x')
                      +2j^{\mu\nu}(x)
               \right\} = 0,
\end{equation}
where we have used the definition for $\bar h^{\mu\nu}(x)$ written in 
(\ref{eq:def bar h}) and the tensor $P_{\rho\sigma,\alpha\beta}$ is
given by
\begin{equation}
   P_{\rho\sigma,\alpha\beta}
        \ = \ {1\over2}\left( \eta_{\rho\alpha}\eta_{\sigma\beta}
                             +\eta_{\rho\beta}\eta_{\sigma\alpha}
                             -\eta_{\rho\sigma}\eta_{\alpha\beta}
                       \right).
\end{equation} 
Note that this differential stochastic equation includes a non-local
term responsible for the dissipation of the gravitational field and a
noise source term which accounts for the fluctuations of the thermal 
plasma. They are connected by a FDR as described in the next section.
Note also that this equation in combination with the correlation for 
the stochastic variable (\ref{eq:correlation}) determine the two-point
correlation for the stochastic metric fluctuations 
$\langle \bar h_{\mu\nu}(x) \bar h_{\alpha\beta}(x') \rangle_j$
self-consistently.


\section{Fluctuation-dissipation relation}
\label{sec:FDR}


Now we want to see how the kernels found above to
represent the noise and the dissipation of the system are functionally
related. This FDR reflects the balance between the quantum
fluctuations of the thermal plasma and the energy loss by the
gravitational field. First, we explicitly see how this
relation appears at zero temperature and compare with the results
obtained in Refs.~\cite{CV96,CV97}. Secondly, we discuss in some
detail the relation in the high temperature limit. Finally, using
the properties of the thermal propagators, we formally show that the
FDR is satisfied in general.

\subsection{Fluctuation-dissipation relation at zero temperature}

Since the thermal functions $n_\beta(|q^o|)$ vanish at zero temperature,
the complete zero temperature contribution to the noise and
dissipation kernels can be directly read from
(\ref{eq:N}) and (\ref{eq:D}) 
\begin{equation}
   \kl{\tilde N}{(o)}{k}
        \ = \  \pi^2\int {d^4q\over(2\pi)^4}\ 
               \delta(q^2)\delta[(k+q)^2]
               \left\{ \theta(k^o+q^o)\theta(-q^o)
                      +\theta(-k^o-q^o)\theta(q^o)
               \right\}\kl{T}{}{q,k},
\end{equation}
\begin{equation}
    \kl{\tilde D}{(o)}{k}
        \ = \  -i\pi^2\int {d^4q\over(2\pi)^4}\ 
               \delta(q^2)\delta[(k+q)^2]
               \left\{ \theta(k^o+q^o)\theta(-q^o)
                      -\theta(-k^o-q^o)\theta(q^o)
               \right\}\kl{T}{}{q,k}.
\end{equation}
The computation of these integrals is very simple because of the
presence of the delta functions. Furthermore, since they are
independent of the temperature the final tensor structure cannot
depend on the four-velocity $u^\mu$ that defines the preferred
co-moving coordinate system of the plasma. In fact, this simpler 
structure allow us to write all the integrals in terms of just only two
integrals
\begin{equation}
   \int {d^4q\over(2\pi)^4}\
   \delta(q^2)\delta[(k+q)^2]
   \left\{ \theta(k^o+q^o)\theta(-q^o)
           \pm\theta(-k^o-q^o)\theta(q^o)
   \right\}
        \ = \ {\theta(-k^2)[\theta(k^o)\pm\theta(-k^o)]\over4(2\pi)^3},
\end{equation}
one corresponding to the noise kernel ($+$ sign) and the other to the
dissipation kernel ($-$ sign). The final result for both kernels can
be written in terms of the tensor $\kl{Q}{}{k}$ defined in 
(\ref{eq:Q tensor}) as
\begin{equation}
   \kl{\tilde N}{(o)}{k}
       \ = \  \left( {\alpha\pi k^4\over4} \right)
                      \theta(-k^2)\kl{Q}{}{k},
\end{equation}
\begin{equation}
    \kl{\tilde D}{(o)}{k}
        \ = \  -i\ sg(k^o)\left( {\alpha\pi k^4\over4} \right)
                          \theta(-k^2)\kl{Q}{}{k}.  
\end{equation}
It is clear from these expressions, that the FDR satisfied by
the noise and dissipation kernel at zero temperature is given by
\begin{equation}
   \kl{\tilde N}{(o)}{k}
        \ = \ -i\ sg(k^o) \kl{\tilde D}{(o)}{k}.
\end{equation}
This zero temperature result includes in one expression the FDR found
in \cite{CV96} for conformally coupled fields and that found in
\cite{CV97} for non-conformally coupled fields.
We next discuss the pure thermal fluctuation contribution in the
leading order at high temperature.

\subsection{Fluctuation-dissipation relation at high temperature}

Here, we describe in some detail the computation of the noise and 
dissipation kernels at high temperature in order to find the
corresponding FDR. The computation of thermal integrals involving
the thermal factors $n_\beta(|q^o|)$ is further complicated by the
tensor structure of gravitation.

At high temperature one considers that the external momentum is much
less than the temperature; that is $k^o, |\vec{k}| \ll \beta^{-1}$.
Here we assume that the temperature is high enough so that this
condition is satisfied, but remains well below the Planck
temperature $\beta^{-1} \ll \beta^{-1}_P \sim \ell_P$ so that graviton
loop corrections to the effective action may be neglected. 
On the other hand, the leading contribution can be
obtained for internal momentum $|\vec{q}|$ of the order of the
temperature $\beta^{-1}$ which means that  the external momentum 
$|\vec{k}|$ is small compared with the internal momentum $|\vec{q}|$;
{\sl i.e.}, $|\vec{k}| \ll |\vec{q}|$ (see, for
example \cite{LeB96}). Let us start with the leading contribution
($\beta^5$) for the noise kernel. The thermal contribution to the 
noise kernel $\kl{\tilde N}{(\beta)}{k}$ is given by the last three 
terms of Eq.~(\ref{eq:N})
\begin{equation}
   \kl{\tilde N}{(\beta)}{k}  
        \ = \ \pi^2\int {d^4q\over(2\pi)^4}\ 
               \delta(q^2)\delta[(k+q)^2]
               \left\{n_\beta(|q^o|)+n_\beta(|k^o+q^o|)
                      +2n_\beta(|q^o|)n_\beta(|k^o+q^o|)
               \right\}\kl{T}{}{q,k}.
\end{equation}
Integrating the $q^o$ variable, and taking into account the properties
of the tensor $\kl{T}{}{q,k}$ the above expression simplifies to
\begin{eqnarray}
   \kl{\tilde N}{(\beta)}{k}  
        & \ = \ & {\pi\over2}\int {d^3\vec{q}\over(2\pi)^3}\
                  {1\over2q^o} {1\over2p^o}
                  [  n_\beta(q^o) + n_\beta(p^o) 
                   +2n_\beta(q^o)n_\beta(p^o)]
                \nonumber\\
        &       & \hskip3cm\times
                  [ \delta(p^o-q^o-k^o)
                   +\delta(p^o+q^o+k^o)] \kl{T}{}{q,k}
                \nonumber\\
        &       &+\{ k \leftrightarrow -k \},
   \label{eq:HTN} 
\end{eqnarray}
where $q^o = |\vec{q}|$ and $p^o = |\vec{k}+\vec{q}|$. On dimensional
grounds it is clear that the leading contribution at high temperature
to the tensor $\kl{T}{}{q,k}$ is given by its first term
$\kl{T}{3}{q,k}$ only. We are also going to assume the following
approximations:
\begin{eqnarray}
   p^o
        & \ \simeq \ &  q^o + \vec{k}\hat q,
                     \nonumber\\
   p^o-q^o-k^o
        & \ \simeq \ & -k^o + \vec{k}\hat q,
                     \nonumber\\ 
   p^o+q^o+k^o
        & \ \simeq \ & 2q^o.       
\end{eqnarray}
Further simplifications arise because in this limit the thermal factor
$n_\beta(p^o)$ may be approximated by
\begin{equation}
   n_\beta(p^o)
        \ \simeq \  n_\beta(q^o) 
                  + \vec{k}\hat q {dn_\beta(q^o)\over dq^o}.
   \label{eq:n approx}
\end{equation}
For the particular combination of thermal functions in the
noise kernel this yields
\begin{equation}
   n_\beta(q^o) + n_\beta(p^o) + 2n_\beta(q^o)n_\beta(p^o)
        \ \simeq \ 2n_\beta(q^o)[1 + n_\beta(q^o)].
\end{equation}
As can be seen by a direct calculation the integrals with 
$\delta(p^o+q^o+k^o)$ do not contribute in this approximation. Then,
the evaluation of the leading contribution at high temperature for the
noise kernel (\ref{eq:HTN}) is reduced to calculate a solid angle
integral and standard thermal integrals, separately,
\begin{equation}
   \kl{\tilde N}{(\beta)}{k}
        \ = \ {\theta(k^2)\over16\pi^2|\vec{k}|}
              \int d\Omega\ \delta(KQ) \kl{T}{3}{Q,K}
              \int^\infty_0 dq^o\ (q^o)^4 
              n_\beta(q^o)[1 + n_\beta(q^o)].
\end{equation}
In order to make the solid angle integral dimensionless we have
used the external momentum $K$ defined previously and we have
introduced a new lightlike vector $Q^\mu \equiv (1,\hat q)$. 
Note that the theta function $\theta(k^2)$ appears as a condition 
imposed by the delta function in the solid angle integral. 
The explicit computation of these last two integrals in terms of the 
tensor basis $\kl{T}{i}{u,K}$ gives the final result to leading order 
in $\beta$
\begin{equation}
   \kl{\tilde N}{(\beta)}{k}
        \ \simeq \  {\pi^3\over30\beta^5}{\theta(k^2)\over|\vec{k}|}
                    \sum^{14}_{i=1}\mbox{\rm N}_i(r)\kl{T}{i}{u,K},
\end{equation}
where the $14$ coefficients $\mbox{\rm N}_i(r)$ are given by
\begin{eqnarray}
   \mbox{\rm N}_i(r)
        & \ = \ & {1\over8}
                  \left[  K^4, 5K^6, 35K^8, K^4, 5K^6, 5K^4r,
                          35K^6r,-(K^2-4r^2)K^2, -5(K^2-6r^2)K^4,
                  \right.
                \nonumber \\
        &       & \hskip.5cm
                  \left. -5(K^2-6r^2)K^4, -5(3K^2-4r^2)K^2r,
                          3K^4-24K^2r^2+8r^4, -(K^2-4r^2)K^2, 5K^4r  
                  \right].
   \label{eq:coeff}
\end{eqnarray}
The determination of these coefficients involves the evaluation of the
inverse of the cumbersome $14\times 14$ matrix
$M_{ij} 
 \equiv \mbox{$\rm T$}_{i\ \mu\nu , \alpha\beta}(u,K)\kl{T}{j}{u,K}$
\cite{Reb91,ABF94}. However, this can be easily performed with an 
algebraic computer program.

The thermal contribution to the dissipation kernel
$\kl{\tilde D}{(\beta)}{k}$ can be obtained similarly. From the
thermal terms of Eq.~(\ref{eq:D}) we have
\begin{equation}
   \kl{\tilde D}{(\beta)}{k}
        \ = \  -i\pi^2\int {d^4q\over(2\pi)^4}\ 
               \delta(q^2)\delta[(k+q)^2]
               \left\{ sg(k^o+q^o) n_\beta(|q^o|)
                      -sg(q^o)n_\beta(|k^o+q^o|)
               \right\}\kl{T}{}{q,k}.
\end{equation}
This integral can be simplified to 
\begin{eqnarray}
   \kl{\tilde D}{(\beta)}{k}  
        & \ = \ & -{i\pi\over2}\int {d^3\vec{q}\over(2\pi)^3}\
                   {1\over2q^o} {1\over2p^o}
                   \left\{ \delta(p^o-q^o-k^o)
                           [ n_\beta(q^o) - n_\beta(p^o) ]
                   \right.
                \nonumber\\
        &       & \hskip3.5cm
                   \left. -\delta(p^o+q^o+k^o) 
                           [ n_\beta(q^o) + n_\beta(p^o) ]
                   \right\}\kl{T}{}{q,k}
                \nonumber\\
        &       & -\{ k \leftrightarrow -k \}.
   \label{eq:HTD} 
\end{eqnarray}
Again the terms with $\delta(p^o+q^o+k^o)$ do not contribute to the
final result. Using (\ref{eq:n approx}) to approximate the difference 
of thermal functions and following the above computation for the noise
kernel, we can write the leading contribution in $\beta$ to the 
dissipation kernel as
\begin{equation}
   \kl{\tilde D}{(\beta)}{k}
        \ \simeq \ -i{\pi^3\over60\beta^4}
                     {\theta(k^2)k^o\over|\vec{k}|}
                     \sum^{14}_{i=1}
                     \mbox{\rm N}_i(r)\kl{T}{i}{u,K},
\end{equation}
where $\mbox{\rm N}_i(r)$ are the coefficients given in 
(\ref{eq:coeff}). It is clear that these leading contributions 
for the noise and dissipation kernels satisfy a FDR at high 
temperature of the form
\begin{equation}
   \kl{\tilde N}{(\beta)}{k}  
        \ \simeq \ i \left( {2\over\beta k^o} \right) 
                     \kl{\tilde D}{(\beta)}{k}. 
\end{equation}
It is important to note that the leading contribution for the noise 
kernel depends on $\beta^{-5}$ whereas that for the dissipation 
kernel depends on $\beta^{-4}$ and this is a direct consequence of 
the FDR. In contrast to the previous zero limit temperature FDR, this 
result may be interpreted to be a consequence of pure thermal
fluctuations. We next show that formally
there is a FDR valid to any order that includes both quantum and thermal
fluctuations.

\subsection{Fluctuation-dissipation relation at finite temperature}

The existence of a FDR at finite temperature between the noise and
dissipation kernels we have identified in the previous sections is very
easy to prove under general conditions. We begin by showing that the 
FDR naively appears at the level of propagators as a direct 
consequence of the KMS relation \cite{KMS} (see Appendix~\ref{app:TFT}). 
Then, using a generalization of this KMS relation, we see how the FDR 
is also satisfied by our noise and dissipation kernels.

To obtain the FDR at the level of propagators we need to introduce the
Schwinger and the Hadamard propagators. These propagators are defined
as the thermal average of the anticommutator
$G(x-x') 
 \equiv -i\langle [\phi(x),\phi(x')] \rangle_\beta$ and the commutator 
$G^{(1)}_\beta (x-x') 
 \equiv -i\langle \{\phi(x),\phi(x')\} \rangle_\beta$, respectively. 
The first represents the linear response of a relativistic system to
an external perturbation and the second the random fluctuations of the
system itself \cite{KMS,FDR}. Since we can write the KMS
condition satisfied by the propagators $G^\beta_{+-}$ and
$G^\beta_{-+}$ in Fourier space as
\begin{equation}
   \tilde G^\beta_{-+}(k)
        \ = \ e^{\beta k^o}\tilde G^\beta_{+-}(k),
   \label{eq:KMS}
\end{equation}
the Fourier transform of both the Schwinger $\tilde G(k)$ and the
Hadamard $\tilde G^{(1)}_\beta(k)$ propagators can be expressed, for
example, in terms of $\tilde G^\beta_{+-}(k)$ alone
\begin{eqnarray}
   \tilde G(k)
        & \ \equiv \ &  \tilde G^\beta_{-+}(k)
                       -\tilde G^\beta_{+-}(k)
          \    =   \    \left( e^{\beta k^o} - 1
                        \right) \tilde G^\beta_{+-}(k),
                     \nonumber\\
   \tilde G^{(1)}_\beta(k)
        & \ \equiv \ &  \tilde G^\beta_{-+}(k)
                       +\tilde G^\beta_{+-}(k)
          \    =   \    \left( e^{\beta k^o} + 1
                        \right) \tilde G^\beta_{+-}(k).
\end{eqnarray}
The FDR satisfied by these propagators follows inmediately from the
above equalities \cite{KMS,FDR}
\begin{equation}
   \tilde G^{(1)}_\beta(k)
        \ = \ \coth \left( {\beta k^o\over2}\right)\tilde G(k).
   \label{eq:FDR propagators} 
\end{equation}
Obviously, this relation can also be recovered if we write the
explicit expressions for the Fourier transform of the propagators
\begin{eqnarray}
   \tilde G(k)
        & \ = \ & -2\pi i\ sg(k^o) \delta(k^2),
                \nonumber\\
   \tilde G^{(1)}_\beta(k)
        & \ = \ & -2\pi i\ 
                   \coth \left({\beta|k^o|\over2 }\right)
                   \delta(k^2).
\end{eqnarray}
To use this last approach in our case could be a very difficult task
because one needs to compute the integrals for the noise and
dissipation kernels explicitly. On the other hand, if we follow the first
technique we only need to generalize the KMS condition of
Eq.~(\ref{eq:KMS}) to product of two propagators.
This generalization reads
\begin{equation}
   \tilde G^\beta_{-+}(k+q)\tilde G^\beta_{+-}(q) 
        \ = \ e^{\beta k^o}
              \tilde G^\beta_{+-}(k+q)\tilde G^\beta_{-+}(q),
\end{equation}
and can be used to deduce the following formal identity
\begin{equation}
   \tilde G^\beta_{-+}(k+q)\tilde G^\beta_{+-}(q)
  +\tilde G^\beta_{+-}(k+q)\tilde G^\beta_{-+}(q)
        \ = \ \coth\left({\beta k^o\over2}\right)
              [ \tilde G^\beta_{-+}(k+q)\tilde G^\beta_{+-}(q)
               -\tilde G^\beta_{+-}(k+q)\tilde G^\beta_{-+}(q)].
   \label{eq:identity}
\end{equation}
Finally, one only has to write, from the trace of Eq.~(\ref{eq:trace})
and the definitions (\ref{eq:N}) and (\ref{eq:D}), the noise and
dissipation kernels in terms of the propagators 
$\tilde G^\beta_{\pm\mp}$, respectively, as
\begin{equation}
   \kl{\tilde N}{}{k}
        \ = \  -{1\over4}\int {d^4q\over(2\pi)^4}\ 
                [ \tilde G^\beta_{-+}(k+q)\tilde G^\beta_{+-}(q)
                 +\tilde G^\beta_{+-}(k+q)\tilde G^\beta_{-+}(q)]
                \kl{T}{}{q,k},
   \label{eq:noise}
\end{equation}
\begin{equation}
    \kl{\tilde D}{}{k}
        \ = \  {i\over4}\int {d^4q\over(2\pi)^4}\ 
               [ \tilde G^\beta_{-+}(k+q)\tilde G^\beta_{+-}(q)
                -\tilde G^\beta_{+-}(k+q)\tilde G^\beta_{-+}(q)]
               \kl{T}{}{q,k},
   \label{eq:dissipation}
\end{equation}
and use the formal equality (\ref{eq:identity}) to prove
that they are related by the thermal identity
\begin{equation}
   \kl{\tilde N}{}{k} 
        \ = \ i\coth\left({\beta k^o\over2}\right)\kl{\tilde D}{}{k}.
\end{equation}
In coordinate space we have the analogous expression
\begin{equation}
   \kl{N}{}{x} 
        \ = \ \int d^4x'\ \mbox{\rm K}_{FD}(x-x')\kl{D}{}{x'},
\end{equation}
where the fluctuation-dissipation kernel $\mbox{\rm K}_{FD}(x-x')$ is
given by the integral
\begin{equation}
   \mbox{\rm K}_{FD}(x-x')
        \ = \ i \int {d^4k\over(2\pi)^4}\
                     e^{ik\cdot(x-x')}
                     \coth\left({\beta k^o\over2}\right).
\end{equation}
The proof of this FDR at finite temperature is in some sense formal
because we have assumed along the argument that the integrals are 
always well defined. Nevertheless, the exact results obtained for the
zero and high temperature limits indicate that the noise and
dissipation kernels are well defined distributions \cite{Jon82}. The
asymptotic analysis has also been useful to determine the physical
origin of the fluctuations.


\section{Linear response theory}
\label{sec:LRT}


The purpose of this section is to understand the connection between the
LRT \cite{CW51,Kub57,KMS,CS77,Mot86} and the functional
methods we have used here. In the spirit of LRT the gravitational field
is considered as a weak external source which imparts
disturbances to the plasma whose response is studied to linear order.

Let us first recall the main features of LRT.
Consider a system described by the Hamiltonian
operator $\hat H_o$ initially coupled to an external 
driving agent linearly, say $A_\alpha$. Since we are only interested
in how the system responds to the external agent, and not the details of
the agent, we will ignore the Hamiltonian for the external perturbation
but write the complete operator Hamiltonian of the sytem as
\begin{equation}
   \hat H
        \ = \ \hat H_o + A_\alpha \hat J^\alpha, 
\end{equation}
where $J^\alpha$ is the current operator associated with the external agent.
If the system is in thermal equilibrium before the external source is
applied the first order response of the system to this external force
is given by the thermal expectation value of the
commutator of the current operator over its thermal average
\begin{equation}
   \langle [J^\mu(x)-\langle J^\mu(x) \rangle_\beta,
            J^\nu(x')-\langle J^\nu(x') \rangle_\beta] 
 \rangle_\beta.
\end{equation}
In contrast, the intrinsic quantum fluctuations of the system are
described by the thermal average of the anticommutator. In our case, 
the conserved current operator is given by the stress-energy tensor 
$T^{\mu\nu}(x)$ as derived from the classical action. The objective of
this final section is to show that the response and fluctuation
functions for the stress-energy tensor considered in the LRT are 
equivalent to our dissipation and noise kernels, respectively.

First, we write the classical action for the matter field to
linear order in the gravitational perturbations,
\begin{equation}
   S_m[\phi,h_{\mu\nu}]
        \ \simeq \ {1\over2}\int d^4x\ 
                   [ \phi\Box\phi + h_{\mu\nu}T^{\mu\nu} ],
\end{equation}
with the stress-energy tensor given by
\begin{equation}
   T^{\mu\nu}
        \ = \ P^{\mu\nu,\alpha\beta} 
              \partial_\alpha\phi\partial_\beta\phi
             +\xi \left( \eta^{\mu\nu}\Box - \partial^\mu\partial^\nu
                  \right) \phi^2.
\end{equation}
Note that $T^{\mu\nu}(x)$ is conserved if the classical unperturbed
equation of motion for $\phi$ is satisfied and it reduces to the 
stress-energy tensor for a scalar field in flat spacetime if 
$\xi = 0$. Alternatively, we can write the Hamiltonian formulation of
our problem. If we introduce the conjugate momentum variable of the
matter field to first order in the perturbation
\begin{equation}
   \Pi  
        \ \equiv \ {\partial {\cal L}\over\partial\dot\phi}
        \ \sim   \ \dot\phi 
                  +{1\over2}h_{\mu\nu}
                   {\partial T^{\mu\nu}\over\partial\dot\phi},
\end{equation}
the Hamiltonian can be written as
\begin{equation}
   H
        \ \simeq \  {1\over2}\int d^3\vec{x}
                    \left[ \Pi^2
                          +(\vec{\nabla}\phi)^2
                          - h_{\mu\nu}T^{\mu\nu}
                    \right]. 
\end{equation}
Note that to first order $\dot\phi$ and $\Pi$ are interchangeable in
the expression for the stress-energy tensor.

Using the thermal version of the Wick theorem \cite{Mil69,LeB96}, one 
can write, after some algebra, the equilibrium thermal average of the
two-point function for the stress-energy tensor operator at different 
spacetime points in terms of products of thermal propagators,
\begin{equation}
   \langle T^{\mu\nu}(x) T^{\alpha\beta}(x') \rangle_\beta
  -\langle T^{\mu\nu}(x) \rangle_\beta
   \langle T^{\alpha\beta}(x') \rangle_\beta
        =   -2\int {d^4k\over(2\pi)^4} e^{ik\cdot (x-x')}
                   \int {d^4q\over(2\pi)^4}\ 
                   \tilde G^\beta_{-+}(k+q)
                   \tilde G^\beta_{+-}(q)
                   \kl{T}{}{q,k}.
\end{equation}
Finally, defining
$ \Delta_\beta T^{\mu\nu}(x) \equiv  T^{\mu\nu}(x) 
 -\langle T^{\mu\nu}(x) \rangle_\beta$
and using the expressions for the the noise and dissipation kernels
given in (\ref{eq:noise}) and (\ref{eq:dissipation}) respectively,
we obtain
\begin{equation}
   \langle \{ \Delta_\beta T^{\mu\nu}(x),
              \Delta_\beta T^{\alpha\beta}(x')
            \} 
   \rangle_\beta 
        \ = \  8\ \kl{N}{}{x-x'},
\end{equation}
\begin{equation}
   \langle [ \Delta_\beta T^{\mu\nu}(x),
              \Delta_\beta T^{\alpha\beta}(x')
            ] 
   \rangle_\beta 
        \ = \  8i\ \kl{D}{}{x-x'}.
\end{equation}
From these formal identities we conclude that the functional method
gives a description of the lowest order dynamics of a near-equilibrium
system equivalent to that given traditionally by the LRT.


\section{Conclusions}
\label{sec:conclusions}


In this paper we show how the functional methods can be used effectively
to study the non-equilibrium dynamics of a thermal quantum field
in an external gravitational field. The close time path (CTP)
effective action and the influcence functional were used
to derive the noise and dissipation kernels of the open system.
We show the formal equivalence of this method with the traditional
linear response theory (LRT) for lowest order perturbance of a
near-equilibrium system, and how the response functions
such as the contribution of the quantum scalar field to the thermal
graviton polarization tensor can be derived.
An important quantity not usually obtained in LRT approaches but
of equal manifest importance in the CTP approach is the
noise term arising from the quantum and thermal fluctuations in the plasma
field. With this we derived a Langevin-type equation
for the non-equilibrium dynamics of the gravitational field under the
influence of the plasma. The back reaction
of the plasma on the gravitational field is embodied in a
fluctuation-dissipation relation (FDR), which connects
the quantum fluctuations of the thermal plasma and the energy
dissipated by the external gravitational field.
We formally prove the existence of such a relation for thermal plasmas
at all temperatures.

This powerful method perfected in recent years 
has been successfully applied to semiclassical gravity and cosmological
problems by a number of authors. It should also prove useful
for thermal field theory problems in QED and QCD, such as electroweak
phase transition, quark-gluon plasma and disoriented chiral condensates
in heavy-ion collision processes.


\acknowledgments


The work of A.C. was supported in part by the CICYT research Project
No. AEN95-0882, the Comissionat per a Universitats i Recerca under a
cooperative agreement between MIT and Generalitat de Catalunya and 
funds provided by the U.S. Department of Energy (D.O.E.) under 
cooperative research agreement DE-FC02-94ER40818. BLH was supported 
in part by NSF grant PHY94-21849.

\vskip1cm

\appendix




\section{Thermal field theory}
\label{app:TFT}


In this Appendix, we describe the CTP formalism in flat spacetime in
order to compute the effective action for a scalar field in thermal 
equilibrium \cite{CSHY85,Mil69,SW85,Paz90}. Using the CTP formalism
one can deal with non-equilibrium situations because we do not have to
assume anything about the final state of the system in contrast to
other formalisms of quantum field theory at finite temperature. 
\cite{NS84,LvW87,LeB96}. We begin with a brief 
general description of the CTP formalism for an arbitrary initial 
state. Then, we explain with more detail the particular features 
of the formalism for an initial state in thermal equilibrium.

\subsection{Arbitrary initial state}

The CTP formalism is a
functional technique to obtain expectation values of products of
quantum field operators \cite{Sch61} and dynamical equations for the
mean field which are real and causal \cite{DeW86,Jor86}. For simplicity,
consider a scalar field $\Phi(x)$ described initially by the time
independent Hamiltonian $H[\Phi(x)]$. Suppose that an external
classical source $J(x)$ is turned on and the perturbed Hamiltonian
assumes the form $H + J(x)\Phi(x)$. If the initial quantum state of
the field is defined by the density matrix operator $\rho$, we can 
define the CTP generating functional, $Z_\rho [J]$, as the following 
statistical average over the initial quantum state
\cite{Kel65,CSHY85,CH88,CHKMP94},
\begin{equation}
   Z_\rho [J] 
        \ \equiv \ Tr\{ \rho\ T_C e^{i \int_C d^nx\ J(x)\phi(x)}\}.
   \label{eq:def gen funct}
\end{equation}
In the above definition $\phi(x)$ is the scalar field operator which 
evolves in time under the unperturbed Hamiltonian $H$, $n$ is the
number of spacetime dimensions, $T_C$ is the time ordering operator 
defined over a contour $C$ in complex time which goes forward in time
and then back to the initial point (see Fig.~1) and the integral 
symbol $\int_C$ means complex integration along this path. Note that 
in this picture the time evolution of the initial state is determined 
by the external source $J(x)$ alone. Our main concern is to 
apply this formalism to a system initially in thermal
equilibrium.
 

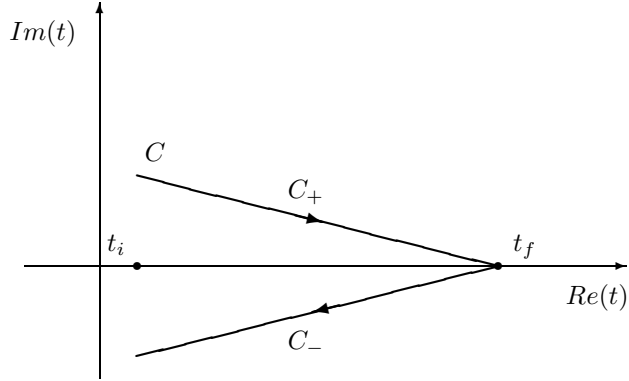
\begin{figure}
\setlength{\unitlength}{1mm}
\vspace{5cm}
\begin{center}
\begin{picture}(100,1)
\put(10,0){\vector(0,1){50}}
\put(0,15){\vector(1,0){80}}
\put(72,10){\makebox(0,0)[bl]{$Re(t)$}}
\put(-2,45){\makebox(0,0)[bl]{$Im(t)$}}
\put(11,17){\makebox(0,0)[bl]{$t_i$}}
\put(65,17){\makebox(0,0)[bl]{$t_f$}}
\put(16,29){\makebox(0,0)[bl]{$C$}}
\put(35,24){\makebox(0,0)[bl]{$C_+$}}
\put(35,4){\makebox(0,0)[bl]{$C_-$}}
\put(15,15){\circle*{1}}
\put(63,15){\circle*{1}}
{\thicklines \put(15,27){\vector(4,-1){24.7}}
             \put(39,21){\line(4,-1){24.1}}
             \put(63,15){\vector(-4,-1){24.7}}
             \put(39,9){\line(-4,-1){24.1}}}
\end{picture}
\end{center}
\caption{Integration contour $C$ in the complex time
         plane for the CTP formalism of non-equilibrium quantum
         fields. This path can be splitted in two separated segments 
         $C_+\cup C_-$, the first going forward in time and the 
         second backward. The small deviation from the real axis 
         is needed to ensure a well defined path integral
         representation of the generating functional.}
\label{fig:CTP contour}
\end{figure}


By differentiating with respect to the external source and then taking 
$J=0$ this functional generates the expectation value over the initial
quantum state specified by the density matrix $\rho$ of time ordered 
products of the field operator along the contour $C$. In particular, 
the two-point function is given by,
\begin{equation}
   {1\over Z_\rho[0]}
   \left. {\delta^2 Z_\rho[J]\over i\delta J(x) i\delta J(x')}
   \right|_{J=0}
        \ = \ {Tr\{ \rho\ T_C \phi(x)\phi(x') \}
               \over Tr\{ \rho  \}}.
\end{equation}
To write an integral representation of the generating functional we have
to introduce a complete basis of eigenstates $\{|\varphi,t_i\rangle\}$
of the field operator $\phi(x)$ at the initial time $t_i$, {\sl i.e.} 
$\phi(t_i,\vec{x})|\varphi,t_i\rangle 
 = \varphi(\vec{x})|\varphi,t_i\rangle$ where $\varphi(\vec{x})$ is
the eigenvalue. This representation for $Z_\rho[J]$ is
\begin{equation}
   Z_\rho[J] 
        \ = \ \int d[\varphi]d[\varphi']\
              \langle\varphi,t_i|\rho|\varphi',t_i\rangle
              \langle\varphi',t_i|T_C e^{i \int_C d^nx\ J(x)\phi(x)}
              |\varphi,t_i\rangle.
\end{equation}
The last term in the integral may be interpreted as a transition 
amplitude from the initial state $|\varphi,t_i\rangle$ to another initial 
state $|\varphi',t_i\rangle$ following the close time path $C$, and
then, it can be expressed as a path integral with the appropriate 
boundary conditions,
\begin{equation}
   Z_\rho[J] 
        \ = \ \int d[\varphi]d[\varphi']\
              \langle\varphi,t_i|\rho|\varphi',t_i\rangle
              \int^{\phi(t_i)=\varphi'}
                  _{\phi(t_i)=\varphi}
              {\cal D}[\phi]\
              e^{i\int_C d^nx\ ({\cal L}[\phi(x)]+J(x)\phi(x))}.
   \label{eq:path int rep}  
\end{equation}
In the above equation, ${\cal L}[\phi(x)]$ is the Lagrangian 
density corresponding to the unperturbed Hamiltonian of the scalar 
field theory. Now, as usual, we introduce the generating functional 
$W_\rho[J] \equiv -i\ \ln Z_\rho[J]$ that generates the connected part
of the $n$-point functions and define the CTP effective action as its 
Legendre transform,
\begin{equation}
   \Gamma^{CTP}_\rho[\bar\phi]
        \ \equiv \ W_\rho[J] - \int_C d^nx\ J(x)\bar\phi(x),
\end{equation}
where $\bar\phi(x)$ is the expectation value of the 
field $\phi(x)$ over the initial state in the presence of the external 
source $J(x)$,
\begin{equation}
   \bar\phi(x)
        \ \equiv \ {\delta W_\rho[J]\over \delta J(x)}
        \ \equiv \ {Tr\{ \rho\ T_C \phi(x)
                         e^{i \int_C d^nx'\ J(x')\phi(x')}
                       \}
                    \over
                    Tr\{ \rho\ T_C
                         e^{i \int_C d^nx'\ J(x')\phi(x')}
                       \}
                   }.
\end{equation}
Note that $\bar\phi$ becomes the mean value of the field if we 
substitute $J=0$. The CTP effective action is the generating 
functional of one-particle irreducible graphs and contains all 
the quantum corrections to the classical action. Since we have only
needed information about the initial state of the system a real 
dynamical equation with causal boundary conditions can be derived
from this effective action . 


\begin{figure}
\setlength{\unitlength}{1mm}
\vspace{5cm}
\begin{center}
\begin{picture}(100,1)
\put(10,0){\vector(0,1){50}}
\put(0,35){\vector(1,0){80}}
\put(72,30){\makebox(0,0)[bl]{$Re(t)$}}
\put(-2,45){\makebox(0,0)[bl]{$Im(t)$}}
\put(11,37){\makebox(0,0)[bl]{$t_i$}}
\put(65,37){\makebox(0,0)[bl]{$t_f$}}
\put(17,10){\makebox(0,0)[bl]{$C_\beta$}}
\put(35,44){\makebox(0,0)[bl]{$C_+$}}
\put(35,24){\makebox(0,0)[bl]{$C_-$}}
\put(2,5){\makebox(0,0)[bl]{$-i\beta$}}
\put(15,35){\circle*{1}}
\put(63,35){\circle*{1}}
\put(10,4){\circle*{1}}
{\thicklines \put(15,47){\vector(4,-1){24.7}}
             \put(39,41){\line(4,-1){24.1}}
             \put(63,35){\vector(-4,-1){24.7}}
             \put(39,29){\line(-4,-1){24.1}}
             \put(15,23){\vector(0,-1){13}}
             \put(15,17){\line(0,-1){13}}}
\end{picture}
\end{center}
\caption{Integration contour in the complex time
         plane for the CTP path integral useful for a 
         system with an initial state in thermal 
         equilibrium.}
\label{fig:finite temperature CTP contour}
\end{figure}
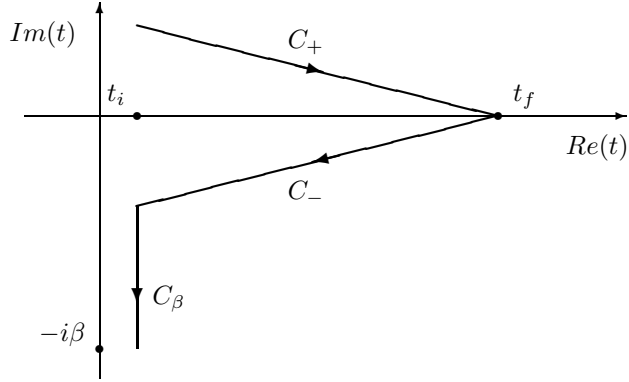


\subsection{Initial state in thermal equilibrium}

For practical purposes, we are going to describe in some detail how to
obtain the Green functions and the effective action for a quantum system
with an initial state in thermal equilibrium at temperature
$\beta^{-1}$ \cite{SW85,CH88,Paz90}. Consider the Lagrangian density
for a free massive scalar field
\begin{equation}
   {\cal L}[\phi(x)]
        \ = \ -{1\over2}\left( \partial_\mu\phi\partial^\mu\phi
                              +m^2\phi^2
                        \right).
   \label{eq:lagrangian}
\end{equation}
The initial thermal state of this system is described by a normalized 
density matrix of the form $\rho_\beta=e^{-\beta H}/Tr\{e^{-\beta H}\}$, 
where $H$ is the Hamiltonian operator corresponding to 
(\ref{eq:lagrangian}) at some initial time $t_i$; {\sl i.e.}, the
unperturbed Hamiltonian of the system before the external source
$J(x)$ is connected. In this case, the path integral representation
of the generating functional given in Eq.~(\ref{eq:path int rep}) can
be written as
\begin{equation}
   Z_\beta[J] 
        \ = \ {1\over Tr\{e^{-\beta H}\}}
              \int d[\varphi]d[\varphi']\
              \langle\varphi,t_i-i\beta|\varphi',t_i\rangle
              \int^{\phi(t_i)=\varphi'}
                  _{\phi(t_i)=\varphi}
              {\cal D}[\phi]\
              e^{i\int_C d^nx\ ({\cal L}[\phi(x)]+J(x)\phi(x))}.  
\end{equation}
In deriving this expression, we have used the fact that the
density matrix operator $\rho_\beta$ can be seen as a time translation
operator in the complex plane because the unperturbed Hamiltonian is the
generator of time evolution for the eigenstates of the field operator
$\phi(x)$. In other words, since the field transforms as 
$\phi(t,\vec{x})=e^{itH}\phi(0,\vec{x})e^{-itH}$ and the eigenstates
evolve in time simply as 
$|\varphi,t\rangle = e^{itH}|\varphi,0\rangle$, the density matrix
generates a translation in complex time $t \rightarrow t - i\beta$. 
The above representation for $Z_\beta[J]$ may be written in a more 
compact form as 
\begin{equation}
   Z_\beta[J] 
        \ = \ {\cal N}\int {\cal D}[\phi]\
              e^{i\int_{C} d^nx\ ({\cal L}[\phi(x)]+J(x)\phi(x))}.  
\end{equation}
The path integral is defined over all possible field configurations 
along the time path $C$ with initial boundary condition 
$\phi(t_i) = \phi(t_i - i\beta)$ and the normalization term ${\cal N}$
includes all the factors which are independent of the external source 
$J(x)$. We would be also able to write the above compact expression 
for the path integral along the time path in the complex plane of 
Fig.~2. But, the contribution to the path integral along the segment
$C_\beta$ has no dynamical consequences in our case because the 
external source is zero along this vertical segment. In the usual 
path integral approach to thermal field theory \cite{NS84,LvW87,LeB96}
(see also \cite{SW85}), the external source is also non zero along 
$C_\beta$ by definition in contrast with our approach 
\cite{CH88,GR94,Paz90}. However, it can be justified that both are 
consistent asymtotically; that is, when 
$t_i \rightarrow -\infty$ and $t_f \rightarrow \infty$. We are going 
to assume this condition from now on.

To perform the path integral of the generating functional it is
convenient to make the change of field variables
$\phi(x) \rightarrow   
         \phi(x) - \int_C d^nx'\ G^C_\beta(x-x')J(x')$,
where the propagator $G^C_\beta(x-x')$ satisfies the differential
equation,
\begin{equation}
   \left( \Box_C - m^2 \right) G^C_\beta(x-x')
        \ = \ \delta_C(x-x').
   \label{eq:dif equ}
\end{equation}
The subscript $C$ in the d'Alambertian $\Box_C$ and the 
delta function $\delta_C$ means that these objects are defined 
along the complex path $C$. The subscript $\beta$ in $G^C_\beta$ is
written to remind that this propagator has to be constructed with the
appropriate thermal boundary conditions. If we redefine the
normalization factor ${\cal N}$ by $Z_\beta[0]$ in order to include a
new path integral independent of $J(x)$, the change of
variable yields
\begin{equation}
   Z_\beta[J]
        \ = \ Z_\beta[0]\
              e^{-{i\over2}\int_C d^nxd^nx'\ 
                 J(x)G^C_\beta(x-x')J(x')}.
\end{equation}
It is clear from the thermal version of (\ref{eq:def gen funct}) and
the above expression for the generating functional that the propagator
$G^C_\beta(x-x')$ can be interpreted as the thermal average of two
time-ordered operators along $C$
\begin{equation}
   G^C_\beta(x-x')
        \ = \ -i \langle T_C \phi(x)\phi(x')
                 \rangle_\beta.
\end{equation}
Above we have introduced the shorthand notation
$\langle A \rangle_\beta \equiv Tr\{ \rho_\beta A \}$
for the trace of a general operator $A$ with respect to the thermal 
density operator. It does not take more additional work at this point 
to show that the Wick theorem also applies at finite temperature 
\cite{Mil69,LeB96}. Let us find an explicit expression for the thermal
propagator. Using the step function $\theta_C$, defined also along the
contour $C$, one can decompose the propagator as
\begin{equation}
   G^C_\beta(\tau-\tau';\vec{x}-\vec{x}')
        \ = \  \theta_C(\tau-\tau')\
               G^+_\beta(\tau-\tau';\vec{x}-\vec{x}')
              +\theta_C(\tau'-\tau)\
               G^-_\beta(\tau-\tau';\vec{x}-\vec{x}'),
\end{equation}
where we have denoted the time variable along the path
$C$ by $\tau$ and defined the propagators
$G^+_\beta(\tau-\tau';\vec{x}-\vec{x}') = 
 G^-_\beta(\tau'-\tau;\vec{x}'-\vec{x}) \equiv
 -i\langle \phi(\tau;\vec{x})\phi(\tau';\vec{x}')
   \rangle_\beta$. The trace nature of the generating functional and
the Hamiltonian time evolution of the field operator $\phi(x)$ impose
some relations on certain propagators. For example, they lead to
equalities like 
$G^-_\beta(\tau;\vec{x}) = G^+_\beta(\tau-i\beta;\vec{x})$ which are
generally known as Kubo-Martin-Schwinger (KMS) relations 
\cite{Kub57,KMS}. In fact, these conditions are a direct consequences
of thermal equilibrium and are necessary to preserve the periodic
boundary condition in the change of field variable previously 
performed in order to simplify the path integral representation of the
generating functional \cite{KK86,Nie89}. The solution of 
(\ref{eq:dif equ}), compatible with the KMS relation, may be expressed
in terms of the thermal function $n_\beta(p^o)=(e^{\beta p^o}-1)^{-1}$
as 
\begin{equation}
   G^C_\beta(\tau-\tau';\vec{x}-\vec{x}')
        \ = \  \int {d^np\over(2\pi)^n}\ e^{ip\cdot (x-x')}
               \left[ \theta_C(\tau-\tau') + n_\beta(p^o)
               \right](-2\pi i)\  sg(p^o)\ \delta(p^2+m^2).
\end{equation}
Now, for simplicity in actual computations, it is convenient to 
introduce a matrix notation for the thermal propagator $G^C_\beta$  
which takes into account the segment of the path, $C_+$ or $C_-$, 
where the time variable is choosen to lie. We have four different
possibilities: 
\begin{itemize}
   \item if $\tau, \tau' \in\ C_+$ or $\tau, \tau' \in\ C_-$ we may
         write, respectively,
         \begin{equation}
            G^\beta_{\pm\pm}(x-x')
                 \ = \  \int {d^np\over(2\pi)^n}\ e^{ip\cdot(x-x')}
                        \left[ {\mp1\over p^2+m^2\mp i\epsilon}
                              -2\pi i\ n_\beta(|p^o|)\delta(p^2+m^2)
                        \right], 
            \label{eq:thermal prop 1}           
         \end{equation}
   \item if $\tau \in\ C_+$ and $\tau' \in\ C_-$, or vice versa, then
         respectively, 
         \begin{equation}
            G^\beta_{\pm\mp}(x-x')
                 \ = \  \int {d^np\over(2\pi)^n}\ e^{ip\cdot(x-x')}
                        (-2\pi i)[\theta(\mp p^o)+n_\beta(|p^o|)] 
                        \delta(p^2+m^2).
            \label{eq:thermal prop 2}
         \end{equation}
\end{itemize}
Since the time variable in the above equations takes values on
the real axis alone, it is very easy to see that all these propagator
components may be expressed as thermal averages
\begin{eqnarray}
   G^\beta_{++}(x-x')
        & \ = \ & -i\langle T \phi(x)\phi(x') \rangle_\beta
                \nonumber \\
   G^\beta_{--}(x-x')
        & \ = \ & -i\langle \bar T \phi(x)\phi(x') \rangle_\beta
                \nonumber \\
   G^\beta_{-+}(x-x') 
        & \ = \ &  -i\langle \phi(x)\phi(x') \rangle_\beta
          \ = \     G^\beta_{+-}(x'-x),
\end{eqnarray}
where $T$ and $\bar T$ are the usual time and anti-time ordering
operators, respectively. If we define in the same way the external
sources $J_\pm(t,\vec{x}) \equiv J(\tau,\vec{x})$ for $\tau\in C_\pm$,
we can write the generating functional in matrix form
\begin{equation}
   Z_\beta[J_a]
        \ = \ Z_\beta[0]\
              e^{-{i\over2}\int d^nxd^nx'\ 
                 J^T_a(x)G^\beta_{ab}(x-x')J_b(x')}.
   \label{eq:matrix form gen funct}
\end{equation}
In the exponents, the time integrals range from $-\infty$ to
$+\infty$, the subindixes $a, b$ take values $+, -$ and the transposed
source vector is defined by $J^T_a(x) \equiv (\ J_+(x), -J_-(x)\ )$. 
The minus sign in the second component of the source vector is a relic
to the fact that originally the integral for the time variable along
the segment $C_-$ was from $+\infty$ to $-\infty$. If we were
interested to specify initial vacuum boundary conditions 
($\rho_o \equiv |0,in\rangle\langle 0,in|$) instead of thermal
equilibrium conditions we could still use 
(\ref{eq:matrix form gen funct}) substituting the thermal propagator 
$G^\beta_{ab}$ by the vacuum propagator 
$G^o_{ab} \equiv \lim_{\beta\rightarrow\infty}G^\beta_{ab}$ (see, for
example \cite{CH87,CV94,Paz90}).

Finally, we are interested to obtain the CTP effective action at finite
temperature in terms of the propagator. Following \cite{CV94}, where
the derivation of the CTP effective action in the case of an initial
pure vacuum state was explained in some detail, but taking into
account the particular boundary conditions for the thermal case, 
we arrive to the expression (see also \cite{Paz90,GR94})
\begin{equation}
   \Gamma^\beta_{CTP}[\bar\phi_a]
        \ = \ S[\bar\phi_a] - {i\over2}Tr\{ \ln G^\beta_{ab} \},
   \label{eq:therm eff act}
\end{equation}
where $S[\bar\phi_a]$ is the classical action for the scalar field
evaluated at the expectation value of $\phi_a(x)$ in the presence of 
the external source $J_a(x)$.

\subsection{CTP effective action for two fields
         in the semiclassical approximation}
\label{app:just}

We now derive the CTP effective action at finite
temperature for two fields in the semiclassical approximation in order
to generalize Eq.~(\ref{eq:therm eff act}) and make it of direct 
applicability to our problem. For simplicity, consider two different
scalar fields described by the general Lagrangian density
\begin{equation}
   {\cal L}[\phi(x)] + {\cal L}[\psi(x),\phi(x)],
\end{equation}
where the second term is assumed to be quadratic in the field
$\psi(x)$. In our case, the first part, ${\cal L}[\phi(x)]$, will
represent the Lagrangian density of the gravitational field, and the
second part, the Lagrangian density of a thermal
scalar field coupled to gravity. Since, we are only interested in the
dynamics of the gravitational field we perturb the system by an
external source coupled only to $\phi(x)$. The CTP generating
functional for this system will be
\begin{equation}
   e^{iW_\beta[J_\pm]} 
        \ \equiv \ \int {\cal D}[\phi]{\cal D}[\psi]\
                   e^{i\int_{C} d^nx\ ( {\cal L}[\phi(x)]
                                       +{\cal L}[\psi(x),\phi(x)]
                                       +J(x)\phi(x))},  
\end{equation}
where $C$ is the time integration path of Fig.~1 and the field
$\psi(x)$ has the appropriate thermal boundary conditions. If one of
the fields is assumed to evolve classically, in our case $\phi(x)$,
the above CTP generating functional may be approximated by
\begin{equation}
   e^{iW_\beta[J_\pm]} 
        \ \simeq \ e^{i\int_{C} d^nx\ ( {\cal L}[\bar\phi(x)]
                                       +J(x)\bar\phi(x))}
                   \int {\cal D}[\psi]\
                   e^{i\int_{C} d^nx\ {\cal L}[\psi(x),\bar\phi(x)]},  
\end{equation}
where $\bar\phi(x)$ is the classical field written in terms of
$J(x)$. The remaining path integral is in fact a functional of 
$\bar\phi(x)$ alone and can be identified with the influence
functional of Feynman and Vernon \cite{FV63}. Since 
${\cal L}[\psi(x),\bar\phi(x)]$ has been assumed to be quadratic in 
$\psi(x)$, the path integral is Gaussian and may be performed
exactly. In matrix notation we have
\begin{equation}
   \exp(-{i\over2}Tr\{\ln\bar G^\beta_{ab}[\bar\phi_\pm]\}),
\end{equation}
where $\bar G^\beta_{ab}[\bar\phi_\pm]$ is the corresponding matrix
propagator with thermal boundary conditions of the Lagrangian density
${\cal L}[\psi(x),\bar\phi(x)]$. When this result is substituted in 
the CTP generating functional and the Legendre transform with respect 
$J(x)$ is done, we can finally write the thermal CTP effective action 
as
\begin{equation}
   \Gamma^\beta_{CTP}[\bar\phi_\pm]
        \ = \ S[\bar\phi_+] 
             -S[\bar\phi_-]
             -{i\over2}Tr\{ \ln\bar G^\beta_{ab}[\bar\phi_\pm]\}.
   \label{eq:twofields}
\end{equation}
This generalized expression for the CTP effective action justifies the
use of Eq.~(\ref{eq:eff act two fields}).



\section{Tensor basis}
\label{app:basis}


The complicated tensor structure of the thermal CTP effective action
may be simplified if the following tensor basis is used

\begin{eqnarray}
   \kl{T}{1}{X,Y}
        & \ = \ & \eta^{\alpha\nu}\eta^{\beta\mu}
                 +\eta^{\alpha\mu}\eta^{\beta\nu}
                \nonumber \\
   \kl{T}{2}{X,Y}
        & \ = \ & X^\mu X^\beta\eta^{\alpha\nu}
                 +X^\mu X^\alpha\eta^{\beta\nu}
                 +X^\nu X^\beta\eta^{\alpha\mu}
                 +X^\nu X^\alpha\eta^{\beta\mu}
                \nonumber \\
   \kl{T}{3}{X,Y}
        & \ = \ & X^\alpha X^\beta X^\mu X^\nu
                \nonumber \\
   \kl{T}{4}{X,Y}
        & \ = \ & \eta^{\mu\nu}\eta^{\alpha\beta}
                \nonumber \\
   \kl{T}{5}{X,Y}
        & \ = \ & X^\mu X^\nu\eta^{\alpha\beta}
                 +X^\alpha X^\beta\eta^{\mu\nu}
                \nonumber \\
   \kl{T}{6}{X,Y}
        & \ = \ & X^\beta (Y^\nu\eta^{\alpha\mu}+Y^\mu\eta^{\alpha\nu})
                 +X^\alpha (Y^\nu\eta^{\beta\mu}+Y^\mu\eta^{\beta\nu})
                \nonumber \\
        &       &+Y^\beta (X^\nu\eta^{\alpha\mu}+X^\mu\eta^{\alpha\nu})
                 +Y^\alpha (X^\nu\eta^{\beta\mu}+X^\mu\eta^{\beta\nu})
                \nonumber \\
   \kl{T}{7}{X,Y}
        & \ = \ & Y^\nu X^\alpha X^\beta X^\mu
                 +Y^\mu X^\alpha X^\beta X^\nu
                 +Y^\beta X^\alpha X^\mu X^\nu
                 +Y^\alpha X^\beta X^\mu X^\nu
                \nonumber \\
   \kl{T}{8}{X,Y}
        & \ = \ & Y^\beta Y^\nu\eta^{\alpha\mu}
                 +Y^\beta Y^\mu\eta^{\alpha\nu}
                 +Y^\alpha Y^\nu\eta^{\beta\mu}
                 +Y^\alpha Y^\mu\eta^{\beta\nu}
                \nonumber \\
   \kl{T}{9}{X,Y}
        & \ = \ & Y^\mu Y^\nu X^\alpha X^\beta
                 +Y^\alpha Y^\beta X^\mu X^\nu
                \nonumber \\
   \kl{T}{10}{X,Y}
        & \ = \ & (Y^\beta X^\alpha + Y^\alpha X^\beta)
                  (Y^\nu X^\mu + Y^\mu X^\nu)
                \nonumber \\
   \kl{T}{11}{X,Y}
        & \ = \ & Y^\beta Y^\mu Y^\nu X^\alpha
                 +Y^\alpha Y^\mu Y^\nu X^\beta
                 +Y^\alpha Y^\beta Y^\nu X^\mu
                 +Y^\alpha Y^\beta Y^\mu X^\nu
                \nonumber \\
   \kl{T}{12}{X,Y}
        & \ = \ & Y^\alpha Y^\beta Y^\mu Y^\nu
                \nonumber \\
   \kl{T}{13}{X,Y}
        & \ = \ & Y^\mu Y^\nu\eta^{\alpha\beta}
                 +Y^\alpha Y^\beta\eta^{\mu\nu}
                \nonumber \\
   \kl{T}{14}{X,Y}
        & \ = \ & (Y^\nu X^\mu + Y^\mu X^\nu)\eta^{\alpha\beta}
                 +(Y^\beta X^\alpha + Y^\alpha X^\beta)\eta^{\mu\nu}.
\end{eqnarray}
This basis of 14 tensors with four indixes is constructed with two 
different four-vectors $X^\mu$ and $Y^\mu$. Note that in order to
maintain the symmetries of the effective action each member of the
basis has to be symmetric under the interchanges 
$\mu\leftrightarrow\nu$, $\alpha\leftrightarrow\beta$
and $(\mu\nu)\leftrightarrow(\alpha\beta)$ \cite{Reb91}.
In terms of this tensor basis, the tensor $\kl{T}{}{q,k}$, introduced
in Sec.~\ref{subsec:CTP eff act}, is defined by 

\begin{eqnarray}
   \kl{T}{}{q,k}
        & \ \equiv \ & \kl{T}{3}{q,k}
                      +\left\{ {1\over4}[(k\cdot q)+q^2]^2
                              +\xi k^2[(k\cdot q)+q^2]
                              +\xi^2 k^4
                       \right\} \kl{T}{4}{q,k}
                     \nonumber \\
        &            &-\left\{  {1\over2}[(k\cdot q)+q^2]
                               +\xi k^2
                       \right\} \kl{T}{5}{q,k}
                      +{1\over2} \kl{T}{7}{q,k}
                      +\xi \kl{T}{9}{q,k}
                     \nonumber \\
        &            &+{1\over4} \kl{T}{10}{q,k}
                      +{1\over2} \xi \kl{T}{11}{q,k}
                      +\xi^2 \kl{T}{12}{q,k}
                     \nonumber \\
        &            &-\xi \left\{ {1\over2}[(k\cdot q)+q^2]
                                  +\xi k^2
                           \right\} \kl{T}{13}{q,k}
                      -{1\over2} \left\{ {1\over2}[(k\cdot q)+q^2]
                                        +\xi k^2
                                 \right\} \kl{T}{14}{q,k}.
\end{eqnarray}



\section{Some coefficients at high temperature}
\label{app:coef}


The $14$ coefficients $\mbox{\rm H}_i(r)$ needed to write the leading
contribution at high temperature of the kernel $\kl{\tilde H}{}{k}$
in terms of the tensor basis of Appendix~B are
\begin{eqnarray}
   \mbox{\rm H}_{1}(r)
        & \ = \ & -{1\over6}
                  -{1\over24}K^2
                  -{1\over8}K^4L
                \nonumber\\
   \mbox{\rm H}_{2}(r)
        & \ = \ & -{1\over3}
                  -{1\over12}K^2
                  -{5\over24}K^4
                  -{5\over8}K^6L
                \nonumber\\
   \mbox{\rm H}_{3}(r)
        & \ = \ & -{1\over3}K^2
                  -{7\over12}K^4
                  -{35\over24}K^6
                  -{35\over8}K^8L
                \nonumber\\
   \mbox{\rm H}_{4}(r)
        & \ = \ & -{1\over24}K^2
                  -{1\over8}K^4L
                \nonumber\\
   \mbox{\rm H}_{5}(r)
        & \ = \ & -{1\over12}K^2
                  -{5\over24}K^4
                  -{5\over8}K^6L
                \nonumber\\
   \mbox{\rm H}_{6}(r)
        & \ = \ & \left( -{1\over12}
                         -{5\over24}K^2
                         -{5\over8}K^4L
                  \right)r
                \nonumber\\
   \mbox{\rm H}_{7}(r)
        & \ = \ & \left( -{1\over3}
                         -{7\over12}K^2
                         -{35\over24}K^4
                         -{35\over8}K^6L 
                  \right)r
                \nonumber\\
   \mbox{\rm H}_{8}(r)
        & \ = \ & -{1\over12}
                  +{5\over24}K^2
                  -{1\over2}K^2L
                  +{5\over8}K^4L
                \nonumber\\
   \mbox{\rm H}_{9}(r)
        & \ = \ & -{1\over6}
                  -{2\over3}K^2
                  +{35\over24}K^4
                  -{15\over4}K^4L
                  +{35\over8}K^6L
                \nonumber\\
   \mbox{\rm H}_{10}(r)
        & \ = \ & -{1\over6}
                  -{2\over3}K^2
                  +{35\over24}K^4
                  -{15\over4}K^4L
                  +{35\over8}K^6L
                \nonumber\\
   \mbox{\rm H}_{11}(r)
        & \ = \ & \left( -{1\over4}
                         +{35\over24}K^2
                         -{5\over2}K^2L
                         +{35\over8}K^4L
                  \right)r
                \nonumber\\
   \mbox{\rm H}_{12}(r)
        & \ = \ & +{13\over12}
                  -{35\over24}K^2
                  - L
                  + 5K^2L
                  -{35\over8}K^4L
                \nonumber\\
   \mbox{\rm H}_{13}(r)
        & \ = \ & -{1\over12}
                  +{5\over24}K^2
                  -{1\over2}K^2L
                  +{5\over8}K^4L
                \nonumber\\
   \mbox{\rm H}_{14}(r)
        & \ = \ & \left( -{1\over12}
                         -{5\over24}K^2
                         -{5\over8}K^4L
                  \right)r
\end{eqnarray}
where $K^\mu \equiv k^\mu/|\vec{k}| \equiv (r,\hat k)$ and the 
function $L$ is given by
\begin{equation}
   L(r)
        \ = \ {r\over2}\ln\left( {r+1\over r-1} \right) -1.
\end{equation}



\end{document}